\documentclass{article}
\usepackage{float}

\def \lket {|}
\def \rket {\rangle}

\newcommand{\ket}[1]{\lket #1\rket}

\newcommand{\comment}[1]{}

\newtheorem{Theorem}{Theorem}
\newtheorem{Lemma}{Lemma}
\newtheorem{Claim}{Claim}
\def\A{{\cal A}}
\def\B{{\cal B}}

\newcommand{\proof}{\noindent {\bf Proof: }}
\newcommand{\qed}{\nobreak \ifvmode \relax \else
      \ifdim\lastskip<1.5em \hskip-\lastskip
      \hskip1.5em plus0em minus0.5em \fi \nobreak
      \vrule height0.75em width0.5em depth0.25em\fi}
\floatstyle{boxed}
\newfloat{Algorithm}{bt}{lop}
\floatname{Algorithm}{Algorithm}

\begin{document}
\title{Quantum search with variable times}
\author{Andris Ambainis\thanks{Institute for Quantum Computing and Department
of Combinatorics and Optimization, University of Waterloo. 
E-mail: {\tt ambainis@math.uwaterloo.ca}. Supported by
NSERC, CIAR, MITACS, ARO and IQC University Professorship.}}

\date{}
\maketitle

\begin{abstract}
Since Grover's seminal work, quantum search has been studied in 
great detail. In the usual search problem, we have a collection of $n$ items
$x_1, \ldots, x_n$ and we would like to find $i: x_i=1$. 
We consider a new variant of this problem 
in which evaluating $x_i$ for different $i$ may take different 
number of time steps. 

Let $t_i$ be the number of time steps required to evaluate $x_i$.
If the numbers $t_i$ are known in advance, we give an algorithm
that solves the problem in $O(\sqrt{t_1^2+t_2^2+\ldots+t_n^2})$ steps.
This is optimal, as we also show a matching lower bound.
The case, when $t_i$ are not known in advance, can be solved
with a polylogarithmic overhead. We also give an application of 
our new search algorithm to computing read-once functions. 
\end{abstract}

\section{Introduction}

Grover's quantum search algorithm \cite{Grover} is one of two most important
quantum algorithms. It allows to search a collection of $n$ items 
in $O(\sqrt{n})$ quantum steps. This gives a
quadratic speedup over the exhaustive search for a variety of search
problems \cite{Ambainis}. 

An implicit assumption is that examining any two items can be 
examined in the same number of time steps. This is not necessarily
true when Grover's algorithm is applied to a specific search problem.
It might be the case that some possible solutions to the search 
problem can be checked faster than others.

Let $t_i$ be the number of time steps required to check the $i^{\rm th}$
solution. Classically, searching for an item $i:x_i=1$ requires
time $\Theta(t_1+\ldots+t_n)$.
A naive application of Grover's search would be to use
$O(\sqrt{n})$ steps, with the maximum possible query time 
$t_{max}=\max_i t_i$ in each step. This gives a
$O(\sqrt{n} t_{max})$ time quantum algorithm.

In this paper, we give a better quantum algorithm. We consider
two settings:
\begin{enumerate}
\item
The times $t_i$ are known in advance and can be used to design 
the algorithm;
\item
The times $t_i$ are not known in advance. The algorithm learns
$t_i$ only if it runs the computation for checking the $i^{\rm th}$ item
for $t_i$ (or more) steps.
\end{enumerate}

For the first setting, we give a quantum algorithm that searches in time
in time $O(\sqrt{T})$ where $T=t^2_1+\ldots+t^2_n$.
For the second, more general setting, we give 
an $O(\sqrt{T}\log^2 T \log^2 \log T)$
time quantum algorithm. We show a lower bound of $\Omega(\sqrt{T})$ 
for the first and, hence, also the second setting.

We give an application of our search algorithm, to 
computing read-once Boolean functions. A Boolean function
$f(x_1, \ldots, x_N)$ is read-once if $f$ has a formula
(consisting of AND, OR and NOT operations) in which every
of the variables $x_1, \ldots, x_N$ appears at most once.
We show that any read-once Boolean function for which the depth of
the read-once formula is $d$ can be computed using 
$O(\sqrt{N} \log^{d-1} N)$ queries. Previously, such algorithm
was only known for the case of balanced 
AND-OR trees \cite{BCW,HMW}.

The model in which queries to different $x_i$ take different time
has been previously studied by H\o yer et al. \cite{HS} who 
proved composition theorems for quantum lower bounds in a similar model. 
Our paper appears to be the first to
study the complexity of quantum search in such model.

\section{Model}

We would like to model the situation when the variable $x_i$ is computed
by an algorithm $\A_i$ which is initialized in the state $\ket{0}$
and, after $t_i$ steps,
outputs the final state $\ket{x_i}\ket{\psi_i}$ for some unknown $\ket{\psi_i}$. 
(For simplicity,
we assume that $\A_i$ always outputs the correct $x_i$.) 
In the first $t_{i}-1$ steps, $\A_i$ can be in arbitrary 
intermediate states.

Our goal is to find $i:x_i=1$. (We sometimes refer to $i:x_i=1$ as
{\em marked items} and $i:x_i=0$ as {\em unmarked}.) Our procedure $\A$ 
can run the algorithms $\A_i$, for some number of steps $t$,
with $\A_i$ outputting $x_i$ if $t_i\leq t$ or ``the computation
is not complete" if $t_i>t$.
The computational cost is the amount of time that is spent running
algorithms $\A_i$. Any transformations that do not involve $\A_i$ are free.
This is a generalization of the usual quantum query model. 

For completeness, we include a more formal definition of our model 
in the appendix. Our algorithms, however, can be understood with
just the informal description in the previous two paragraphs.

{\bf Known vs. unknown times.}
We consider two variants of this model. In the ``known times" model, 
the times $t_1, \ldots, t_n$ are known in advance and can 
be used to design the algorithm. In the ``unknown times" model,
$t_1, \ldots, t_n$ are unknown to the designer of the algorithm.

\section{Methods and subroutines}

\subsection{Amplitude amplification}
\label{sec:aa}

Amplitude amplification \cite{BHMT} is a generalization of Grover's quantum
search algorithm.
Let 
\begin{equation}
\label{eq:amplify1} 
\sin \alpha \ket{1}\ket{\psi_1}+\cos\alpha\ket{0}\ket{\psi_0} 
\end{equation}
be the final state of a quantum algorithm $\A$ that outputs 1 with probability
$\sin^2\alpha=\delta$. We would like to increase the probability of the 
algorithm outputting 1. Brassard et al. \cite{BHMT} showed that, by 
repeating $\A$ and $\A^{-1}$ $2m+1$ times, it is possible to generate the final state
\begin{equation}
\label{eq:amplify2} 
\sin (2m+1) \alpha \ket{1}\ket{\psi_1}+\cos (2m+1) \alpha\ket{0}\ket{\psi_0} .
\end{equation}
In particular, taking $m=O(\frac{1}{\sqrt{\delta}})$ 
achieves a constant probability of answer 1.

We use a result by Aaronson and Ambainis \cite{AA} 
who gave a tighter analysis of the same algorithm:
\begin{Lemma}
\label{lem:AA}
\cite{AA}
Let $\A$ be a quantum algorithm that outputs a correct answer and a witness
with probability\footnote{\cite{AA} requires the probability to be exactly
$\epsilon$ but the proof works without changes if the probability is less than
the given $\epsilon$.} $\delta\leq \epsilon$ where $\epsilon$ is known.
Furthermore, let 
\begin{equation}
\label{eq:mconstraint} 
m\leq \frac{\pi}{4 \arcsin \sqrt{\epsilon}} - \frac{1}{2} .
\end{equation}
Then, there is an algorithm $\A'$ which uses $2m+1$ calls to $\A$ and $\A^{-1}$
and outputs a correct answer and a witness with probability 
\begin{equation}
\label{eq:aa} 
\delta_{new}\geq \left( 1-\frac{(2m+1)^2}{3} \delta \right) (2m+1)^2 \delta .
\end{equation}
\end{Lemma}  

The distinction between this lemma and the standard amplitude
amplification is as follows. The standard amplitude amplification 
increases the probability from $\delta$ to $\Omega(1)$ 
in $2m+1=O(\frac{1}{\sqrt{\delta}})$ 
repetitions. In other words, $2m+1$ repetitions increase the success
probability $\Omega((2m+1)^2)$ times. Lemma \ref{lem:AA} achieves an
increase of almost $(2m+1)^2$ times, without the big-$\Omega$ factor.
This is useful if we have an algorithm with $k$ levels of amplitude
amplification nested one inside another. 
Then, with the usual amplitude amplification,
a big-$\Omega$ constant of $c$ would result in a $c^{k}$ 
factor in the running time. Using Lemma \ref{lem:AA} avoids that.

We also need another fact about amplitude 
amplification.

\begin{Claim}
\label{claim:d}
Let $\delta$ and $\delta'$ be such that $\delta\leq \epsilon$ and
$\delta'\leq \epsilon$ and let $m$ satisfy the constraint (\ref{eq:mconstraint}).
Let $p(\delta)$ be the success probability obtained by applying
the procedure of Lemma \ref{lem:AA} to an algorithm with success 
probability $\delta$. If $\delta'\leq \delta\leq c\delta'$ for $c\geq 1$, then
$p(\delta')\leq p(\delta) \leq c p(\delta')$.
\end{Claim}

\proof
Because of equations (\ref{eq:amplify1}), (\ref{eq:amplify2}),
\[ p(\delta)= \sin^2 ((2m+1) \arcsin \sqrt{\delta}) .\]
Let $\gamma=\arcsin \sqrt{\delta}$ and $\gamma'=\arcsin \sqrt{\delta'}$.
Then, we have to prove that $\sin^2 \gamma' \leq \sin^2 \gamma\leq c \sin^2 \gamma'$
implies $\sin^2 (2m+1) \gamma' \leq \sin^2 (2m+1) \gamma\leq c \sin^2 (2m+1) \gamma'$.

Because of $\delta\leq \epsilon$ and $\delta'\leq \epsilon$, we have
$\sqrt{\delta}\leq \sqrt{\epsilon}$ and $\sqrt{\delta'}\leq \sqrt{\epsilon}$.
Together with (\ref{eq:mconstraint}), that means that 
$(2m+1)\arcsin \sqrt{\delta}\leq \frac{\pi}{2}$ and
$(2m+1)\arcsin \sqrt{\delta'}\leq \frac{\pi}{2}$.
Since $\sin$ is an increasing function on $[0, \frac{\pi}{2}]$,
$\sin^2 \gamma' \leq \sin^2 \gamma$
implies $\sin^2 (2m+1) \gamma' \leq \sin^2 (2m+1) \gamma$.

To prove the other inequality, 
we consider the function $f(x)=\frac{\sin (x \gamma)}{\sin (x \gamma')}$.
It suffices to show that $f(x)$ is non-increasing on $[1, 2m+1]$.
(That implies $\frac{\sin^2 ((2m+1)\gamma)}{\sin^2 ((2m+1)\gamma')} \leq
\frac{\sin^2 \gamma}{\sin^2 \gamma'} = c$.)
We have
\[ f'(x)=\frac{\cos (x \gamma) \sin (x\gamma')- \sin (x\gamma) \cos (x\gamma')}{
\sin^2 (x\gamma')} = \frac{\sin x(\gamma'-\gamma)}{\sin^2 (x \gamma')} .\]
This is non-positive, as long as $x(\gamma'-\gamma)\in [\frac{-\pi}{2}, 0]$.
This is true, because 
$x(\gamma'-\gamma)\geq -x\gamma \geq -(2m+1)\gamma\geq -\frac{\pi}{2}$
and $x(\gamma'-\gamma)\leq 0$ follows from $\gamma'\leq \gamma$
(which follows from $\delta'\leq \delta$).
\qed

\comment{The second fact is that, if we have an algorithm $\A$ 
with a success probability $\delta<p$ (where $p\leq \frac{1}{9}$), 
then we can always select $m$ so that the success probability is increased to 
$\delta_{new}$ with $p\leq \delta_{new} \leq 9 p$. To prove that, let $m$
be the largest integer such that the amplitude amplification with $2m+1$ uses of $\A$ 
gives a success probability $\delta'<p$. Let $\delta''$ be the success probability of 
the amplitude amplification with $2m+3$ uses of $\A$. Then, 
\[ \frac{\delta''}{\delta'}=\frac{\sin^2 ((2m+3) \arcsin \sqrt{\delta})}{
\sin^2 ((2m+1) \arcsin \sqrt{\delta})} .\]
By concavity of the $\sin$ function on the interval $[0, \frac{\pi}{2}]$, this is
at most
\[ \frac{(2m+3)^2}{(2m+1)^2} \leq 9.\]
Therefore, $p\leq \delta''< 9p$.

}
\subsection{Amplitude estimation}

The second result that we use is 
a version of quantum amplitude estimation.

\begin{Theorem}
\cite{BHMT}
\label{thm:bhmt}
There is a procedure {\bf Est-Amp}$(\A, M)$ which, given 
a quantum algorithm $\A$
and a number $M$, outputs an estimate $\tilde{\epsilon}$ of
the probability $\epsilon$ that $\A$ outputs 1 
and, with probability at least $\frac{8}{\pi^2}$, we have
\[ |\epsilon-\tilde{\epsilon} | \leq 2 \pi 
\frac{\sqrt{\max(\epsilon(1-\epsilon), 
\tilde{\epsilon}(1-\tilde{\epsilon}))}}{M}+\frac{\pi^2}{M^2} .\]
The algorithm uses $M$ evaluations of $\A$.
\end{Theorem}

We are interested in a slightly different type of error bound.
We would like to have $|\epsilon-\tilde{\epsilon}|\leq 
c \tilde{\epsilon}$ for some small $c>0$. 

\begin{Theorem}
\label{thm:est}
There is a procedure {\bf Estimate}$(\A, c, p, k)$ which, given 
a constant $c$, $0<c\leq 1$ and a quantum algorithm $\A$
(with the promise that the probability $\epsilon$ that 
the algorithm $\A$ outputs 1 is either 0 or at least a 
given value $p$)
outputs an estimate $\tilde{\epsilon}$ of
the probability $\epsilon$ such that, 
with probability at least $1-\frac{1}{2^k}$, 
we have
\begin{enumerate}
\item[(i)]
$|\epsilon-\tilde{\epsilon}|<c \tilde{\epsilon}$ if $\epsilon\geq p$;
\item[(ii)] 
$\tilde{\epsilon}=0$ if $\epsilon=0$.
\end{enumerate}
The procedure {\bf Estimate}$(\A, c, p, k)$ uses the expected number of
\[ \Theta\left(k\left(1+\log \log \frac{1}{p}\right) 
\sqrt{\frac{1}{\max(\epsilon, p)}} \right) \]
evaluations of $\A$.
\end{Theorem}

\proof
We can increase the success probability of {\bf Est-Amp}$(\A, M)$ 
to at least $1-\frac{1}{2^k \log M_{max}}$ (where
$M_{max}=\frac{8 \pi}{c\sqrt{(1-c)p}}$), by repeating
the algorithm $t=O((1+\log \log \frac{1}{p}) k)$ times
and taking the median of the results. 

The procedure {\bf Estimate} calls the repeated {\bf Est-Amp} 
at most $\log M_{max}$ times.
Since each call of {\bf Est-Amp} produces the correct answer
with probability at least $1-\frac{1}{2^k \log M_{max}}$,
the probability that all calls to {\bf Est-Amp} produce correct results
is at least $1-\frac{1}{2^k}$. In this case,
{\bf Estimate} is always correct, because by Theorem \ref{thm:bhmt},
the error $|\tilde{\epsilon}-\epsilon|$ is at most
$\frac{2\pi\sqrt{\epsilon}}{M}+\frac{\pi^2}{M^2}$ and
{\bf Estimate} only stops when this quantity becomes less than
$c\tilde{\epsilon}$. It remains to bound the number of times
{\bf Estimate} calls $\A$.

\begin{Algorithm}
\begin{enumerate}
\item
Let $M=2$;
\item
Repeat:
\begin{enumerate}
\item
Let $\tilde{\epsilon}$ be the estimate output by repeated
{\bf Est-Amp}$(\A, M)$.
\item
\label{step:est-stop}
If $2 \pi \frac{\sqrt{\tilde{\epsilon}(1-\tilde{\epsilon})}}{M}
+\frac{\pi^2}{M^2} \leq c \tilde{\epsilon}$, stop
and output $\tilde{\epsilon}$ as the estimate.
\item
$M=2*M$.
\end{enumerate}
until $M>M_{max}$ where $M_{max}=\frac{8 \pi}{c\sqrt{(1-c)p}}$.
\end{enumerate}
\caption{Procedure {\bf Estimate}}
\end{Algorithm}

If $M\geq \frac{4\pi}{c\sqrt{(1-c)\epsilon}}$, then,
\begin{equation}
\label{eq:eps} 
\frac{2\pi\sqrt{(1-c)\epsilon}}{M}
+2\frac{\pi^2}{M^2} \leq \frac{c(1-c)\epsilon}{2}+
\frac{c^2(1-c)\epsilon}{16}\leq
c (1-c) \epsilon.
\end{equation}
Then, $\tilde{\epsilon}\geq (1-c(1-c))\epsilon\geq 
(1-c)\epsilon$.
Therefore, the quantity of equation (\ref{eq:eps}) 
is less than or equal to $c\tilde{\epsilon}$.
Hence, if $M\geq \frac{4\pi}{c\sqrt{(1-c)\epsilon}}$, then 
the condition in step \ref{step:est-stop} is satisfied and 
the algorithm stops.
Since $M$ is doubled in every iteration,
the final value of $M$ is $M_{0}<\frac{8\pi}{c\sqrt{
(1-c)\epsilon}}$. The algorithm $\A$ is repeated
\[ M_{0}t+\frac{M_{0}t}{2}+\frac{M_0t}{4}+\ldots < 2 M_0 t 
< \frac{16 \pi}{c\sqrt{(1-c)\epsilon}} t \] times.

If $\epsilon\geq p$, the algorithm must stop with $M$ being at
most $\frac{8\pi}{c\sqrt{(1-c)p}}$. If that does not
happen, we can conclude that $\epsilon=0$. The number of
repetitions of $\A$ in this case is at most 
$\frac{16 \pi}{c\sqrt{(1-c)p}} t$.
\qed

\section{Search algorithm: known running times}
\label{sec:known}

\begin{Theorem}
\label{MainThm}
A collection of $n$ items with times $t_1, \ldots, t_n$ 
can be searched in time
\[ O\left( \sqrt{t_1^2+t_2^2+\ldots+t_n^2} \right) .\]
\end{Theorem}

\proof
The basic idea is to subdivide the items into groups so that all items in one group
have similar times $t_i$ (e.g. $\frac{t_{max}}{2} \leq t_i \leq t_{max}$
for some $t_{max}$). We can perform the standard Grover search
in a group in time $s=O(\sqrt{l} t_{max})$ where $l$ is the size of the group.
We then observe that
\[ s^2 = O(l t^2_{max}) = O \left(\sum_i t^2_i \right) ,\]
with the summation over all items $i$ in the same group.
By summing over all groups, we get 
\[ \sum_j s_j^2 = O\left(\sum_{i=1}^N t^2_i\right) ,\]
where $j$ on the left ranges over all groups.
Let $k$ be the number of the groups that we have. 
If we have a search algorithm that searches $k$ items 
in time 
\[ O\left( \sqrt{s_1^2+\ldots+s_k^2} \right) ,\]
we can then substitute the algorithms for searching the $k$ groups instead of
the $k$ items and obtain a search algorithm for $n$ items that runs in
time
\[ O\left( \sqrt{t_1^2+\ldots+t_n^2} \right) .\]
We then design a search algorithm for $k$ items in a similar way.

The simplest implementation of this strategy gives an algorithm with $\log^* n$
levels of recursion and running time 
\[ O\left( c^{\log^* n} \sqrt{t_1^2+t_2^2+\ldots+t_n^2} \right) ,\]
due to the reduction from $n$ items to $k$ items losing a constant
factor every time it is used.
The $c^{\log^* n}$ factor can be avoided, by a more sophisticated
implementation of the same idea, which we describe below.

We first restrict to the case when there is exactly one marked item.
The general case can be reduced to this case with a constant factor overhead,
by running the algorithm on all $n$ elements, a random set of $\frac{n}{2}$,
a random set of $\frac{n}{4}$, etc. 
As shown in \cite{AA}, there is a constant probability that at least 
one of those sets contains exactly one marked item.
The expected running time increases by at most a constant factor,
because of the following lemma.

\begin{Lemma}
Let $S$ be a uniformly random set of $\frac{n}{2^j}$ elements of 
$\{1, 2, \ldots, n\}$. 
Then, 
\[ E \left[ \sqrt{\sum_{i\in S} t^2_i} \right] \leq 
\frac{1}{2^{j/2}}  \sqrt{\sum_{i\in \{1, \ldots, n\}} t^2_i} .\]
\end{Lemma}

\proof
By concavity of the square root function,
\[ E \left[ \sqrt{\sum_{i\in S} t^2_i} \right] \leq 
\sqrt{E\left[\sum_{i\in S} t^2_i\right]} =
 \frac{1}{2^{j/2}}  \sqrt{\sum_{i\in \{1, \ldots, n\}} t^2_i} .\]
\qed

Therefore, the reduction from the general case to one marked item case 
increases the bound on the number of queries by a factor of
at most
\[ 1+\frac{1}{2^{1/2}}+\frac{1}{2}+\ldots < \frac{1}{1-\frac{1}{\sqrt{2}}} .\]

Second, we introduce a generalization of the problem in which 
the algorithm $\A_i$ for the marked $i$ returns the correct answer with 
a probability at least $p_i$, instead of a certainty. More formally,
\begin{itemize}
\item
if $x_i=0$, the final state of the algorithm $\A_i$ is of the form
$\ket{0}\ket{\psi_0}$.
\item 
if $x_i=1$, the final state of the algorithm $\A_i$ is of the form
$\alpha\ket{1}\ket{\psi_1}+\sqrt{1-\alpha^2} \ket{0}\ket{\psi_0}$,
where $p_i \leq |\alpha|^2\leq d \cdot p_i$, for some constant $d>1$.
\end{itemize}
The probabilities $p_1, \ldots, p_n$ and the constant $d$
are known to us when we design 
the algorithm, just as the times $t_1, \ldots, t_n$. 
(Knowing both the success probability and the running time may look quite artificial. 
However, we only use the "known success probability" model
to design an algorithm for the case when all $\A_i$ return 
the correct answer with certainty.)

We claim that, in this case, we can search in time
\[ O\left( \sqrt{\frac{t_1^2}{p_1}+\frac{t_2^2}{p_2}+
\ldots+\frac{t_n^2}{p_n}} \right) .\]
Our main theorem now follows as the particular case $p_1=\ldots=p_n=1$.
The main part of our proof is

\begin{Lemma}
\label{MainLemma1}
There exists $k=O( \log^3 n \log \log n)$ with the following property.
Assume that there is a search algorithm for $k$ items with some fixed $d>1$ 
that works in time at most 
\[ C\sqrt{\frac{s_1^2}{q_1}+\frac{s_2^2}{q_2}+\ldots+\frac{s_k^2}{q_k}}. \]
for any given times $s_1, \ldots, s_k$ and probabilities $q_1, \ldots, q_k$.
Then, there exists a search algorithm for $n$ items with
$d'=\left(1-O\left(\frac{1}{\log n}\right)\right) d$ instead of $d$ that  
works in time at most
\[ C \left(1+O\left(\frac{1}{\log n}\right)\right) 
\sqrt{\frac{t^2_1}{p_1}+\frac{t^2_2}{p_2}+\ldots+\frac{t^2_n}{p_n}} \]
for any given times $t_1, \ldots, t_n$ and probabilities $p_1, \ldots, p_n$.
\end{Lemma}

\proof
Let $T_0$ be the maximum of $\frac{t_1}{\sqrt{p_1}}, \ldots, 
\frac{t_n}{\sqrt{p_n}}$.
We first check all items with $\frac{t_i}{\sqrt{p_i}}\leq \frac{T_0}{n\log n}$ 
sequentially. To check item $i$, we just amplify the success probability of 
$\A_i$ to $\Omega(1)$ by the standard amplitude amplification,
in $O(\sqrt{p_i})$ steps. Therefore, the number of steps for checking  
item $i$ is 
\[ O\left(\sqrt{p_i} \frac{t_i}{p_i}\right)=
O\left(\frac{t_i}{\sqrt{p_i}}\right)=
O\left(\frac{T_0}{n\log n}\right) .\]

The time for checking all such items is at most the number of such items times  
$\frac{T_0}{n \log n}$ which is of the order at most
\[ \frac{T_0}{\log n} \leq \frac{1}{\log n} \sqrt{T_0^2} 
\leq \frac{1}{\log n} \sqrt{\frac{t^2_1}{p_1}+\ldots+\frac{t^2_n}{p_n}}.\]

Next, if $p_i< \frac{1}{9\log n}$, we choose $m$ so that 
$\frac{1}{9\log n} \leq (2m+1)^2 p_i \leq \frac{1}{\log n}$. 
(Such choice of $m$ always exists, because, if 
$(2m+1)^2 p_i < \frac{1}{9\log n}$, then
\[ (2m+3)^2 p_i < \frac{(2m+3)^2}{(2m+1)^2} \frac{1}{9\log n}
\leq 9 \frac{1}{9\log n} \leq \frac{1}{\log n} .\]
Therefore, it suffices to choose the smallest $m$ for
which $(2m+1)^2 p_i\geq \frac{1}{9\log n}$.)
We then apply Lemma \ref{lem:AA}. 
If the success probability is $p_i$, it increases the success
probability to $p'_i$,
while increasing the running time $(2m+1)$ times.
By Claim \ref{claim:d}, if the success probability is between
$p_i$ and $d\cdot p_i$, it increases to a probability between
$p'_i$ and $d\cdot p'_i$.
By Lemma \ref{lem:AA}, the ratio $\frac{t^2_i}{p_i}$ increases at most
\[ \frac{1}{1-\frac{(2m+1)^2}{3} p_i} \leq \frac{1}{1- \frac{1}{3 \log n}}
= 1 + O\left( \frac{1}{\log n} \right) \]
times. 
After that, we have

\begin{Claim}
\label{claim:pT}
\begin{enumerate}
\item[(a)]
$p_0 \leq p_i \leq 1$, where $p_0=\frac{1-o(1)}{9\log n}$.
\item[(b)]
$\frac{T_0\sqrt{p_0}}{n \log n}\leq t_i\leq T_0$.
\end{enumerate}
\end{Claim}

\proof
(a) Let $p'_i$ be the value of $p_i$ before the amplification. 
If $p'_i\geq \frac{1}{9\log n}$, then $p_i=p'_i$.
If $p'_i<\frac{1}{9 \log n}$, then
\[ p_i = \left( 1-\frac{(2m+1)^2 p'_i}{3} \right) (2m+1)^2 p'_i \geq
\left( 1-\frac{1}{3\log n}\right) \frac{1}{9 \log n} .\]

(b) The first inequality follows from 
$\frac{t_i}{\sqrt{p_i}}\geq \frac{T_0}{n \log n}$ and $p_i\geq p_0$.
The second inequality follows from 
$T_0=\max \frac{t_i}{\sqrt{p_i}}$ and $p_i\leq 1$.
\qed

We partition the intervals $[\frac{T_0 \sqrt{p_0}}{n \log n}, T_0]$
and $[p_0, 1]$ into subintervals
of the form $[T', T'']$ and $[p', p'']$ with
$T''\leq (1+\frac{1}{\log n}) T'$ and 
$p''\leq (1+\frac{1}{\log n}) p'$.
It suffices to have $O(\log^2 n)$ intervals for time
and $O(\log n \log \log n)$ intervals for the success probability.
The overall number of intervals is $m=O(\log^3 n \log \log n)$.

We define new search algorithms $A'_j$, where $j$ ranges over
the pairs of intervals $[T', T'']$ and $[p', p'']$.
Let $[T'_j, T''_j]$ and $[p'_j, p''_j]$ be the time and probability
intervals for $A'_j$ and let $S_j$ be the set of $i$ such that 
$T'_j < t_i \leq T''_j$ and $p'_j < p_i \leq p''_j$.
The algorithm $\A'_j$ picks $i\in S_j$ uniformly at random and 
then runs $\A_i$. 

Let $s_j$ denote the running time of $\A'_j$. Then,
$s_j \leq T''_j$. The success probability of $\A'_j$
is in the interval $[q_j, d(1+\frac{1}{\log n}) q_j]$
where $q_j = \frac{p'_j}{|S_j|}$.
We now relate $s_j$ and $q_j$ to $T_j$ and $p_j$:
\[ \frac{(s_j)^2}{q_j} \leq |S_j| \frac{(T''_j)^2}{p'_j} ,\]
\[ \sum_{i \in S_j} \frac{t_i^2}{p_i} \geq |S_j| \frac{(T'_j)^2}{p''_j} =
\left( 1+ O\left(\frac{1}{\log n}\right)\right) 
|S_j| \frac{(T''_j)^2}{p'_j} \geq 
\left( 1+ O\left(\frac{1}{\log n}\right)\right) 
\frac{(s_j)^2}{q_j}.\]
By summing over all pairs of intervals $j$, 
\[ \frac{s_1^2}{q_1}+\ldots+\frac{s_k^2}{q_k} \leq
\left( 1+ O\left(\frac{1}{\log n}\right)\right)
\left( \frac{t_1^2}{p_1}+\ldots+\frac{t_n^2}{p_n} \right).\]
We now apply the search algorithm for $k$ items
to $\A'_1, \ldots, \A'_k$.
\qed

To obtain Theorem \ref{MainThm}, we repeatedly apply Lemma \ref{MainLemma1}
until the number of items becomes less than some constant $n_0$. 
That happens after $O(\log^* n)$ applications of Lemma \ref{MainLemma1}.

Let $t_1, \ldots, t_{n}$ and $p_1$, $\ldots$, $p_n$ be the times
and probabilities for the final $n\leq n_0$ items. 
After that, we just amplify the success probability 
of every item to $\Omega(1)$
(which increases each $\frac{t^2_i}{p_i}$ by at most a constant
factor, as discussed in the proof of Lemma \ref{MainLemma1}). 
We then search $n$ items in time $O(\sqrt{n} \max_i t_i)$,
using the amplitude amplification, with $\max_i t_i$ steps 
for evaluating any of the items $i$.
Since $p_i=\Omega(1)$ and $n\leq n_0$ where $n_0$ is a constant, we have
\[  \sqrt{n} \max t_i = O(\max t_i) = 
O\left(\sqrt{t^2_1+\ldots+t^2_{n}}\right) =
O\left(\sqrt{\frac{t^2_1}{p_1}+\ldots+\frac{t^2_n}{p_{n}}}\right) .\]
$O(\log^* n)$ applications of Lemma \ref{MainLemma1}
increase the time by a factor of
at most $(1+O(\frac{1}{\log n}))^{\log^* n}=1+o(1)$.
\qed

\section{Application: read-once functions}

A Boolean function $f(x_1, \ldots, x_N)$ that depends on all variables
$x_1, \ldots, x_N$ is read-once if
it has a Boolean formula (consisting of ANDs, ORs and NOTs)
in which every variable appears exactly once. A read-once 
function can be represented by a tree in which every leaf 
contains $x_i$ or $\mbox{NOT~} x_i$ and every internal vertex contains 
AND or OR. 

Barnum and Saks \cite{BS} have shown that, for any read-once $f$,
$\Omega(\sqrt{N})$ queries are necessary to compute $f$ in the 
quantum query model. This bound is known to be tight for a special 
class of read-once functions: balanced AND-OR trees. A balanced AND-OR tree is a
read once function represented by a depth-$d$ tree in which each
internal node has $\sqrt[d]{N}$ children. Nodes on the even levels 
are AND nodes, nodes on the odd levels are OR nodes (or opposite).
Hoyer, Mosca and de Wolf \cite{HMW} have shown that, for any constant $d$,
the function corresponding to the AND-OR tree of depth $d$ can be
evaluated with $O(\sqrt{N})$ queries.
This improved over an earlier $O(\sqrt{N}\log^{d-1} N)$ query
algorithm by Buhrman, Cleve and Wigderson \cite{BCW}.
Both of those algorithms depend on the fact that every node on the
same level of the tree has an equal number of children.

We give the first quantum algorithm for the general case,
when the number of children may vary for different nodes at the same 
depth. 

\begin{Theorem}
Any read-once function $f(x_1, \ldots, x_N)$ of depth $d$ can be computed
by a quantum algorithm that uses $O(\sqrt{N} \log^{d-1} N)$ queries.
\end{Theorem}

\proof
By induction. The base case, $d=1$ is just the OR (or AND) function which can be computed
with $O(\sqrt{N})$ queries using Grover's search to search for $i:x_i=1$ (or $i:x_i=0$).

For the inductive case, assume that $f$ is represented by a depth-$d$ tree
with OR at the root. (The case when the root contains AND is similar.)
Let $n$ be the number of vertices on the level 1 (that is, the number of children of the root
vertex) and $t_i$ be the number of vertices in the subtree rooted
in the $i^{\rm th}$ level-1 vertex. By re-ordering the variables,
we can assume that
\[ f(x_1, \ldots, x_N)= \vee_{i=1}^n f_i(x_{t_1+\ldots+t_{i-1}+1}, \ldots, 
x_{t_1+\ldots+t_i}) .\]
To compute $f$, we have to determine if there exists $i\in\{1, \ldots, n\}$
for which $f_i=1$.
By the inductive assumption, there is an algorithm that computes $f_i$
using $O(\sqrt{t_i} \log^{d-2} t_i)=O(\sqrt{t_i} \log^{d-2} N)$ queries.
We repeat this algorithm $O(\log N)$ times to increase the probability
of correct answer to at least $1-\frac{1}{N^2}$. 
Let $\A_i$ be the resulting algorithm and 
$T_i=O(\sqrt{t_i} \log^{d-1} N)$ be the number of queries in $\A_i$.

We now apply Theorem \ref{MainThm} to $\A_1$, $\ldots$, $\A_n$.
This gives an algorithm which uses
\[ O\left(\sqrt{T_1^2+T_2^2+\ldots+T_n^2}\right)= O(\log^{d-1} N \sqrt{t_1+t_2+\ldots+t_n})
= O(\sqrt{N} \log^{d-1} N) \]
queries. Since we are applying Theorem \ref{MainThm}
to $\A_1$, $\ldots$, $\A_n$ which are incorrect with a small probability,
we have to bound the error probability for the resulting algorithm. 

Let $\A'_1$, $\ldots$, $\A'_n$ be the ``ideal versions" of $\A_1$, $\ldots$, $\A_n$.
If the final state of $\A_i$ is 
\begin{equation}
\label{eq:real} 
\alpha \ket{a}\ket{\psi_a}+\sqrt{1-\alpha^2} \ket{1-a}\ket{\psi_{1-a}}, 
\end{equation}
where $a$ is the correct answer (the value of $f_i$), then the final state
of $\A'_i$ is $\ket{a}\ket{\psi_a}$. ($\A'_i$ can be obtained by composing
$\A_i$ with a transformation that maps the state (\ref{eq:real}) to 
$\ket{a}\ket{\psi_a}$.)

Given the ``ideal algorithms" $\A'_i$, the algorithm of Theorem \ref{MainThm}
would output the correct answer with a constant probability (e.g., at least 2/3).
Since each $\A_i$ outputs the correct answer with probability at least 
$1-\frac{1}{N^2}$, replacing $\A_i$ by $\A'_i$ in one time step
changes the state of the algorithm by at most $O(\frac{1}{N})$ 
(in the $l_2$ norm). Replacing $\A_i$ by $\A'_i$ in every time
step changes the state by at most 
\[ O\left(\frac{\sqrt{N} \log^{d-1} N}{N}\right) =
O\left( \frac{\log^{d-1} N}{\sqrt{N}} \right) \]
in $l_2$ norm. Therefore, the success probability will still
be $\frac{2}{3}-o(1)$, even if the actual $\A_1, \ldots, \A_n$ are used. 
\qed

\section{Search algorithm: unknown running times}
\label{sec:unknown}

\begin{Algorithm}
\begin{enumerate}
\item
Set $j=1$. Define $\B_1$ as the algorithm that 
just outputs 1 and a uniformly random $i\in\{1, \ldots, n\}$.
\item
Repeat:
\begin{enumerate}
\item
\label{step:generate}
Use the algorithm $\B_j$ to generate $k=2 \log (D(j +1))$ 
samples $i_1, \ldots, i_k$ 
of uniformly random elements $i\in S_j$. 
Run $2^{j+1}$ steps of the query procedure on each of $i_1, \ldots, i_k$.
If $x_i=1$ for one of samples, output $i$ and stop.
\item
Let $\B'_{j+1}$ be an algorithm that runs $\B_{j}$ once and, if the output bit is 1, takes
the output index $i$ and runs $2^{j+1}$ steps of the checking procedure on $i$.
If the result is $x_i=0$, $\B'_j$ outputs 0. Otherwise, it outputs 1 
and the same index $i$.
\item
\label{step:estimate}
Let $p=\mbox{{\bf Estimate}}(\B'_{j+1}, c, \frac{1}{N}, 
2 \log (D(j+1)))$.
If $p=0$, output ``no $i:x_i=0$''.
\item
\label{step:noamplify}
If $p\geq \frac{1}{9 \log n}$, let $\B_{j+1}$ be $\B'_{j+1}$.
\item
\label{step:amplify}
If $p < \frac{1}{9 \log n}$, let $\B_{j+1}$ be the algorithm 
obtained by amplifying $\B'_{j+1}$ $2m+1$ times, where $m$ is the 
smallest number for which 
$\frac{1}{9\log n}\leq (2m+1)^2 p \leq \frac{1}{\log n}$.
(Such choice of $m$ always exists, as described in the proof
of Lemma \ref{MainLemma1}.)
\item
Let $j=j+1$.
\end{enumerate}
\end{enumerate}
\caption{Search algorithm for unknown $t_1, \ldots, t_n$}
\label{alg:main}
\end{Algorithm}

In some applications, it may be the case that the times $t_i$ are not known in
advance. We can also solve this case, with a polylogarithmic overhead.

\begin{Theorem}
\label{UnknownThm}
Let $\epsilon>0$.
There is an algorithm that searches
collection of $n$ items with unknown times $t_1, \ldots, t_n$ 
and, with probability at least $1-\epsilon$, stops after
\[ O\left( T \log^2 T \log^2 \log  T  \right) \]
steps, where $T=\sqrt{t_1^2+t_2^2+\ldots+t_n^2}$.
\end{Theorem}

\proof
Again, we assume that there is exactly one marked item.
(The reduction from the general case to the one marked item
case is similar to one in the proof of Theorem \ref{MainThm}.)

Let $S_t$ be the set of items such that 
$x_i=1$ or $t_i\geq 2^t$ and let $n_t=|S_t|$.
Our main procedure, algorithm \ref{alg:main},
defines a sequence of algorithms $\B_1$, $\ldots$, $\B_l$.
The algorithm $\B_j$, with some success probability, 
outputs a bit 1 and, conditional on output bit 1, 
it also outputs a uniformly random index $i\in S_{j}$.
To avoid the problem with accumulating constant factors
(described after Lemma \ref{lem:AA}), we make the success
probability of $\B_j$ slightly less than 1.

\begin{Lemma}
\label{lem:correct}
Assume that the constant $D$ in steps \ref{step:generate} and 
\ref{step:estimate} satisfies $D\leq \frac{\pi}{\sqrt{3\epsilon}}$.
Then, with probability $1-\epsilon$, 
the following conditions are satisfied:
\begin{enumerate}
\item[(a)]
Estimates $p$ are accurate within an multiplicative factor of $(1+c)$;
\item[(b)]
If $\B_j$ is defined, then $t_i>2^{j-1}$ for at least
$\frac{n_{j-1}}{2}$ values $i\in\{1, \ldots, n\}$. 
\end{enumerate}
\end{Lemma}

\proof
(a) The probability of error for ${\bf Estimate}$ is 
at most $\frac{1}{D^2(j+1)^2}$. By summing over all $j$, the probability
of error for some $j$ is at most 
\[ \frac{1}{D^2} \sum_{i=1}^{\infty} \frac{1}{i^2} 
= \frac{1}{D^2} \frac{\pi^2}{6} ,\]
which can be made less than $\frac{\epsilon}{2}$ by choosing 
$D\leq \frac{\pi}{\sqrt{3\epsilon}}$.

(b) By definition, $S_{j-1}$ is the set of all $i$ with the property that
either $x_i=1$ or $t_i>2^{j-1}$. Let $S$ be the set of $i$ with $x_i=1$ and  
$t_i\leq 2^{j-1}$. If $|S|\leq \frac{1}{2} n_{j-1}$, (c) is true.
Otherwise, the probability that each $i_j$ generated in step 
\ref{step:generate} does not belong
to $S$ is less than $\frac{1}{2}$. If one of them belongs to $S$,
algorithm \ref{alg:main} stops without defining $\B_j$.
The probability that this does not happen (i.e., all $i_j$
do not belong to $S$) is less than $(\frac{1}{2})^k= \frac{1}{D^2 (j+1)^2}$.
We can make this probability arbitrarily small similarly to part (a).
\qed

We now bound the running time of algorithm \ref{alg:main}, 
under the asumption
that both conditions of Lemma \ref{lem:correct} are true.
For that, we first bound the running time of the algorithms $\B_j$
and then the total running time of algorithm \ref{alg:main}.
We assume that both conditions of Lemma \ref{lem:correct} are true.

Let $p_j$ be the success probability of $\B_j$ and $p'_{j}$ be
the success probability of $\B'_j$.
Let $r_{k, l}$ be the number of times step \ref{step:amplify} is
performed, for $j\in\{k, k+1, \ldots, l-1\}$.

\begin{Lemma}
\label{lem:aj}
The running time of $\B_j$ is at most
\begin{equation}
\label{eq:an} 
\left(1+\frac{C}{\log n}\right)^{r_{1, j}}
\sqrt{\frac{p_j n}{n_j}}+
\sum_{j'=2}^j \left(1+\frac{C}{\log n}\right)^{r_{j', j}}
\sqrt{\frac{p_j n_{j'-1}}{p_{j'-1} n_j}}  2^{j'}  
\end{equation}
for some constant $C$.
\end{Lemma}

\proof
By induction. The base case is easy. Since $p_j=1$ and $n_j=n$,
the expression (\ref{eq:an}) is just equal to 1, which is also
the running time of $\B_1$.

For the inductive case, we first consider the running time of $\B'_{j+1}$.
It can be decomposed into two parts: the running time of 
$\B_j$ and the running time of the $2^{j+1}$-step checking procedure.
The running time of $\B_j$ is described by equation (\ref{eq:an}).
We have $p'_{j+1}=\frac{p_j n_{j+1}}{n_{j}}$. Therefore, we can rewrite
(\ref{eq:an}) as 
\[
\left(1+\frac{C}{\log n}\right)^{r_{1, j}}
\sqrt{\frac{p'_{j+1} n}{n_{j+1}}}+
\sum_{j'=2}^j \left(1+\frac{C}{\log n}\right)^{r_{j', j}}
\sqrt{\frac{p'_{j+1} n_{j'-1}}{p_{j'-1} n_{j+1}}}  2^{j'}  .
\]
The time for the checking procedure is just $2^{j+1}$
which is equal to $\frac{p'_{j+1} n_{j}}{p_{j} n_{j+1}} 2^{j+1}$
(since $\frac{p'_{j+1} n_{j}}{p_{j} n_{j+1}}=1$).
Therefore, the running time of $\B'_{j+1}$ is
\begin{equation}
\label{eq:an1}
\left(1+\frac{C}{\log n}\right)^{r_{1, j}}
\sqrt{\frac{p'_{j+1} n}{n_{j+1}}}+
\sum_{j'=2}^{j+1} \left(1+\frac{C}{\log n}\right)^{r_{j', j}}
\sqrt{\frac{p'_{j+1} n_{j'-1}}{p_{j'-1} n_{j+1}}}  2^{j'}  .
\end{equation}
If step \ref{step:noamplify} is performed, then $\B_{j+1}=\B'_{j+1}$, 
$p_{j+1}=p'_{j+1}$, $r_{j', j}=r_{j', j+1}$
and the expression (\ref{eq:an1}) is the same as (\ref{eq:an}) 
with $j+1$ instead of $j$.

If the step \ref{step:amplify} is performed, the running time 
of $\B_{j+1}$ is $(2m+1)$ times the running time of $\B'_{j+1}$.
The success probability is 
\[ p_{j+1}\geq \left( 1-\frac{(2m+1)^2}{3} p'_{j+1} \right) 
(2m+1)^2 p'_{j+1} \geq
\left(1-\frac{1}{3\log n}\right) (2m+1)^2 p'_{j+1} .\]
Therefore,
\begin{equation}
\label{eq:mbound} 
2m+1 \leq \left(1+\frac{C}{\log n}\right) \sqrt{\frac{p_{j+1}}{p'_{j+1}}} 
\end{equation}
for some constant $C$.
Multiplying (\ref{eq:an1}) by $2m+1$ and applying (\ref{eq:mbound}) completes
the induction step.
\qed

\begin{Lemma}
\label{lem:rbound}
For all $j, j'$, $r_{j, j'}=O(\log n)$.
\end{Lemma}

\proof
We consider the ratio $q_j=\frac{p_j}{n_j}$. 
We have $q_1=\frac{1}{n}$ and $q_j\leq 1$ for all $j$ (since
$p_j\leq 1$ and $n_j\geq 1$).

Next, we relate $q_j$ and $q_{j+1}$. We have 
$\frac{p'_{j+1}}{n_{j+1}}=\frac{p_j}{n_j}$.
If step \ref{step:noamplify} is applied, 
$p_{j+1}=p'_{j+1}$ and $q_{j+1}=\frac{p'_{j+1}}{n_{j+1}}=q_j$.
If step \ref{step:amplify} is applied,
\[ p_{j+1}\geq (2m+1)^2 \left(1-\frac{1}{3 \log n} \right) p'_{j+1} \geq 
9 \left(1-\frac{1}{3 \log n} \right) p'_{j+1} .\]
Therefore, $q_{j+1}\geq 9 (1-\frac{1}{3 \log n}) q_{j}$.
This means that $q_{j'} \geq (9-3\log n)^{r_{j, j'}} q_j$.
Together with $q_{j'}\leq 1$ and $q_j\geq q_1\geq \frac{1}{n}$,
this implies $r_{j, j'}=O(\log n)$.
\qed

The expression of Lemma \ref{lem:aj} can be upper-bounded by

\begin{Lemma}
\label{lem:aj1}
The running time of $\B_j$ is 
\[ O\left( j \sqrt{\log n} \sqrt{\frac{t_1^2+t_2^2+\ldots+t_n^2}{n_j}} \right) .\]
\end{Lemma}

\proof
We look at each of the components of the sum (\ref{eq:an}) separately.
Consider a term
\begin{equation}
\label{eq:term} 
\left(1+\frac{c}{\log n}\right)^{r_{j', j}}
\sqrt{\frac{p_j n_{j'-1}}{p_{j'-1} n_j}}  2^{j'}  .
\end{equation}
Because of Lemma \ref{lem:rbound},
the first multiplier is bounded from above by a constant.
Since $p_{j'-1}\geq \frac{1-o(1)}{9 \log n}$ 
(similarly to Claim \ref{claim:pT}),
we can upperbound (\ref{eq:term}) by 
$O(\sqrt{\log n \frac{n_{j'-1}}{n_j}} 2^{j'})$.
Let $k$ be the number of $i\in\{1, \ldots, n\}$ for which 
$t_i\geq 2^{j'-1}$. By Lemma \ref{lem:correct}, $k\geq\frac{n_{j'-1}}{2}$ and
\[ t_1^2+t_2^2+\ldots+t_n^2 \geq k 2^{j'-1} \geq n_{j'-1} 2^{j'-2} .\]
This means that each term in (\ref{eq:term}) is at most
\[ O\left( \sqrt{\log n} \sqrt{\frac{t_1^2+t_2^2+\ldots+t_n^2}{n_j}} \right) .\]
The lemma follows by summing over all $j$ terms in (\ref{eq:an}). 
\qed

We now bound the overall running time.
To generate a sample from $S_{j}$, 
one needs $O(\sqrt{\log n})$ invocations of $\B_j$
(because the success probability of $\B_j$ is of the order 
$\Omega(\frac{1}{\log n})$).
Therefore, we need $O(\sqrt{\log n} \log j)$ invocations to generate 
$O(\log j)$ samples in step \ref{step:generate}.
By Lemma \ref{lem:aj1}, that can be done in time 
\[ O \left( j \log j \log n \sqrt{\frac{t_1^2+t_2^2+\ldots+t_n^2}{n_j}}
\right) .\] 
For each of those samples, we run the checking procedure with $2^{j+1}$ steps.
That takes at most twice the time required by $\B_j$ (because $\B_j$ includes
the checking procedure with $2^j$ steps). Therefore, the time for the $2^{j+1}$
checking procedure is of the same order or less than the time to generate the samples.

Second, the success
probability estimated in the last step is of order 
$\frac{p_j n_{j+1}}{n_{j}}=\Omega(\frac{n_{j+1}}{n_j \log n})$.
By Lemma \ref{thm:est}, it can be estimated with
\[ O\left(  \log j \log \log n \sqrt{\frac{n_j \log n}{n_{j+1}}}\right) \]
invocations of $\B_j$, each of which runs in time described by
Lemma \ref{lem:aj1}.

Thus, the overall number of steps in one loop of algorithm \ref{alg:main}
is of order at most
\[ \sqrt{t_1^2+t_2^2+\ldots+t_n^2} \left( \frac{j \log j \log n}{\sqrt{n_j}} + 
\frac{j \log j \log n \log \log n }{\sqrt{n_{j+1}}} \right) .\]
Since $n_j\geq 1$ and $n_{j+1}\geq 1$, this is of order
\[ O\left( \sqrt{t_1^2+t_2^2+\ldots+t_n^2} j 
 \log j \log n \log \log n  \right).\]
Let $t_{max}$ be the maximum of $t_1$, $\ldots$, $t_n$.
Then, the maximum value of $j$ is at most $\lceil \log (t_{max}+1) \rceil$.
Therefore, the number of steps used by the algorithm \ref{alg:main}
is 
\[ O\left( \sqrt{t_1^2+t_2^2+\ldots+t_n^2} \log n \log \log n \log t_{max} 
\log \log t_{max} \right).\]
The theorem now follows from $n\leq \sqrt{T}$
and $t_{max} \leq \sqrt{T}$, where $T=t_1^2+t_2^2+\ldots+t_n^2$.
\qed

\section{Search lower bound}

\begin{Theorem}
For any positive integers $t_1, \ldots, t_n$, 
searching a collection of $n$ items
that can be checked in times $t_1, \ldots, t_n$
requires time $c \sqrt{t^2_1+t^2_2+\ldots+t_n^2}$, for
some constant $c>0$.
\end{Theorem}

\proof
Let $t'_i$ be the maximum integer such that
$\lceil \frac{\pi}{4} \sqrt{t'_i} \rceil+1 \leq t_i$
(with $t'_i=1$ if the maximum integer is 0).
We consider searching $m=t'_1+\ldots+t'_n$ elements
$x_1, \ldots, x_m\in\{0, 1\}$ 
in the standard model (where every query 
takes 1 step), with the promise that there is 
either 0 or 1 element $j:x_j=1$. 
By lower bound on quantum search, $c' \sqrt{m}$ 
queries are required to distinguish between the case
when there are 0 elements $j:x_j=1$ and the case
when there is 1 element $j:x_j=1$, for some constant $c'$.

We subdivide the inputs $x_1, \ldots, x_m$
into $n$ groups $S_1$, $\ldots$, $S_n$, with $t'_1, \ldots, t'_n$ 
elements, respectively. Let $y_i=1$ if 
there exists $j\in S_i$ with $x_j=1$.
Since there is either 0 or 1 element $j:x_j=1$,
we know that there is either 0 or 1 element $i:y_i=1$.
We will show

\begin{Lemma}
\label{claim:subroutine}
There is an algorithm that implements the transformation
$\ket{i}\rightarrow \ket{i}\ket{y_i}\ket{\psi_i}$ for some
states $\ket{\psi_i}$, using $t_i$ queries.
\end{Lemma}

Let $\A$ be a search algorithm for search among $n$ items 
that require times $t_1, \ldots, t_n$ and let
$t'$ be the number of steps used by $\A$.
Then, we can substitute the algorithm of Lemma 
\ref{claim:subroutine} instead of the queries $y_i$.
Then, we obtain an algorithm $\A'$ that, 
given $x_1, \ldots, x_n$, asks $t'$ queries and distinguishes
whether there is exactly 1 item $i:y_i=1$ (and, hence, 1 item
$j:x_j=1$) or there is no items $i:y_i=0$ (and, hence, no items
$j:x_j=1$). Hence,
\[ t' \geq c'\sqrt{n}= c' \sqrt{t'_1+\ldots+t'_n} .\]
We now bound $t'_i$ in terms of $t_i$. By definition of $t'_i$, we have
\[ t_i\leq \left\lceil \frac{\pi}{4} \sqrt{t'_i} \right\rceil +1 \leq
\frac{\pi}{4} \sqrt{t'_i} + 2 .\]
This means that $t'_i \geq \frac{16}{\pi^2} (t_i-2)^2$.
If $t_i\geq 3$, then $t_i-2\geq \frac{t_i}{3}$
and $t'_i \geq \frac{16}{9\pi^2} t^2_i$.
If $t_i<3$, then $t'_1\geq 1 \geq  
\frac{16}{9\pi^2} t^2_i$. 
Therefore, 
\[ t' \geq c' \sqrt{t'_1+\ldots+t'_n} \geq 
c' \sqrt{\frac{16}{9\pi^2} (t^2_1+\ldots+t^2_n)} 
= \frac{4c'}{3\pi} \sqrt{t^2_1+\ldots+t^2_n}  .\]
This means that the theorem is true, with $c=\frac{4c'}{3\pi}$.
It remains to prove Lemma \ref{claim:subroutine}.

\proof
[of Lemma \ref{claim:subroutine}]
To simplify the notation, we assume that the group $S_i$ consists
of variables $x_1, \ldots, x_{t_i}$. 
If $t_i=1$, then $y_i=x_1$ and we can just query $x_1$.
This produces the required transformation 
$\ket{i}\rightarrow \ket{i}\ket{y_i}$.

For the $t_i>1$ case,
we have to search $t_i$ items $x_1, \ldots, x_{t_i}$ for an item
$j:x_j=1$, if we are promised that there is either 0 or 1 such item.
There is a modification of Grover's algorithm 
which succeeds with probability 1,
using at most $\lceil \frac{\pi}{4} \sqrt{t_i} \rceil$ queries 
\cite{BHMT}.

The result of Grover's algorithm is:
\begin{itemize}
\item
the state $\ket{j}$, where $j$ is the index for which $x_j=1$, if such $j$ exists;
\item
the superposition $\frac{1}{\sqrt{t_i}} \sum_{j=1}^{t_i} \ket{j}$, otherwise.
\end{itemize}
With one more query (which queries the index $j$), we can determine the value $y_i=x_j$
(which is 1 in the first case and 0 in the second case).
\qed

\section{Conclusion}

In this paper, we gave a quantum algorithm for the generalization of 
Grover's search in which checking different items requires different
times. Our algorithm is optimal for the case when times $t_i$ are known
in advance and nearly optimal (within a polylogarithmic factor) for the
general case. We also gave an application of our algorithm to computing
read-once Boolean functions. It is likely that our algorithms will find
other applications.

While we have mostly resolved the complexity of
search in this setting, the complexity of other problems have not been
studied at all. Of particular interest are problems which are frequently
used as a subroutines in other quantum algorithms (for such
problems, there is a higher chance that the variable-time query version
will be useful). Besides the usual quantum search, the two most
common quantum subroutines are quantum counting \cite{Counting}
and $k$-item search (a version of search in which one has to find $k$ different
$i$ for which $x_i=1$). Element distinctness \cite{Ambainis04,B+} has also
been used as a subroutine, to design quantum algorithms
for the triangle problem \cite{MSS} and verifying matrix identities \cite{BS06,MN}.

{\bf Acknowledgments.}
I would like to thank Robert \v Spalek 
and Ronald de Wolf for the discussion that
lead to this paper.

\begin{appendix}

\section{Formal definition of our model}

To define our model formally, 
let $\A^{(j)}_i$ be the $j^{\rm th}$ step of $\A_i$.
Then, 
\[ \A_i=\A^{(t_i)}_i \A^{(t_i-1)}_i \ldots \A^{(1)}_i .\]
We define $\A^{(t)}_i=I$ for $t>t_i$.
We regard the state space of $\A_i$ as consisting of two registers,
one of which stores the answer ($c\in\{0, 1, 2\}$, with 2 representing 
a computation that has not been completed) and the other register, $x$, 
stores any other information. 

The state space of a search algorithm is spanned by basis states
of the form $\ket{i, t, t_r, c, x, z}$ where $i\in\{1, \ldots, n\}$,
$t, t_r\in\{0, 1, \ldots, T\}$ (with $T$ being the number of the
query steps in the algorithm), $c\in\{0, 1, 2\}$ and $x$ and $z$ 
range over arbitrary finite sets. $i$ represents the index being queried,
$t$ represents the number of the time step in which
the query for $x_i$ started and $t_r$ is the number of time steps for
which $\A$ will run the query algorithm $\A_i$.
$c$ is the output register of $\A_i$ and $x$ holds intermediate 
data of $\A_i$. Both of those registers should be 
initialized to $\ket{0}$ 
at the beginning of every computation of a new $x_i$.
$z$ contains any data that is not a part of the current query.

We define a quantum query algorithm $\A$ 
as a tuple $(U_0, \ldots, U_T)$ of unitary transformations 
that do not depend on $x_1, \ldots, x_n$. 
The actual sequence of transformations that is applied is
\[ U_0, Q_1, U_1, Q_2, \ldots, U_{T-1}, Q_{T}, U_{T}, \]
where $Q_j$ are queries which are defined below.
This sequence of transformations is 
applied to a fixed starting state $\ket{\psi_{start}}$, which
consists of basis states 
$\ket{i, 0, 0, c, x, z}$. 

Queries $Q_j$ are defined in a following way. 
If $j\leq t+t_r$, we apply 
$A^{(j-t)}_i$ to $\ket{c}$ and $\ket{x}$ registers. 
Otherwise, we apply $I$. We call the resulting sequence of
queries $Q_1$, $Q_2$, $\ldots$ {\em generated} by 
transformations $A^{j}_i$.
We call $Q_1$, $Q_2$ a {\em valid} sequence of queries corresponding
to $x_1, \ldots, x_n$ if it is generated by 
$A^{j}_i$ satisfying the following constraints:
\begin{enumerate}
\item
For $t<t_i$, $A^t_i A^{t-1}_i \ldots A^1_i\ket{0}$ is of the
form $\ket{2}\ket{\psi}$ for some $\ket{\psi}$. 
\item
For $t=t_i$, $A^t_i A^{t-1}_i \ldots A^1_i\ket{0}$ is of the
form $\ket{x_i}\ket{\psi}$ for some $\ket{\psi}$. 
\end{enumerate}

$U_j$ can be arbitrary transformations 
that do not depend on $x_1, \ldots, x_n$.

An algorithm $(U_0, \ldots, U_T)$ with the starting state
$\ket{\psi_{start}}$ computes a function $f(x_1, \ldots, x_n)$ if,
for every $x_1, \ldots, x_n\in\{0, 1\}$ and every valid query
sequence $Q_1$, $\ldots$, $Q_T$ corresponding to $x_1, \ldots, x_n$,
the probability of obtaining $f(x_1, \ldots, x_n)$ when
measuring the first qubit of 
\[ U_T Q_T U_{T-1} \ldots U_1 Q_T U_0 \ket{\psi_{start}} \]
is at least 2/3.
\end{appendix}


\begin{thebibliography}{99}

\bibitem{AA}
S. Aaronson, A. Ambainis,
Quantum search of spatial regions.
{\em Theory of Computing},
1:47-79, 2005.
Also quant-ph/0303041.

\bibitem{Ambainis04}
A. Ambainis. Quantum walk algorithm for element distinctness. 
{\em Proceedings of FOCS'04}, pp. 22-31.
Also quant-ph/0311001.

\bibitem{Ambainis}
A. Ambainis. 
Quantum search algorithms. 
{\em SIGACT News}, 35 (2004):22-35. 
Also quant-ph/0504012.

\bibitem{BS}
H. Barnum, M. Saks,
A lower bound on the quantum complexity of read once functions.
{\em Journal of Computer and System Sciences},
69:244-258, 2004.

\bibitem{B+}
H. Buhrman, C. Durr, M. Heiligman, P. H\o yer, F. Magniez, M. Santha, 
R. de Wolf. Quantum algorithms for element distinctness. 
{\em SIAM Journal on Computing}, 34(6): 1324-1330, 2005.
Also quant-ph/0007016.

\bibitem{BS06}
H. Buhrman, R. \v Spalek: Quantum verification of matrix products. 
{\em Proceedings of SODA'06}, pp. 880-889.
Also quant-ph/0409035.

\bibitem{BHMT}
G. Brassard, P. H\o yer, M. Mosca, A. Tapp.
Quantum amplitude amplification and estimation.
In {\em Quantum Computation and Quantum Information Science},
AMS Contemporary Mathematics Series, 305:53-74, 2002.
Also quant-ph/0005055.

\bibitem{Counting}
G. Brassard, P. H\o yer, A. Tapp. 
Quantum counting. {\em Proceedings of ICALP'98}, 
pp. 820-831,
quant-ph/9805082.

\bibitem{BCW}
H. Buhrman, R. Cleve, A. Wigderson,
Quantum vs. classical communication and computation.
{\em Proceedings of STOC'98}, pages 63-68,
quant-ph/9702040.

\bibitem{Grover}
L. Grover.
A fast quantum mechanical algorithm for database search. 
{\em Proceedings of STOC'96}, pp. 212-219.

\bibitem{HMW}
P. H\o yer, M. Mosca, and R. de Wolf. 
Quantum search on bounded-error inputs. 
{\em Proceedings of ICALP'03}, Lecture Notes in Computer Science, 2719:291-299. 
Also quant-ph/0304052 

\bibitem{HS}
P. H\o yer, T. Lee, R. \v Spalek.
Tight adversary bounds for composite functions,
quant-ph/0509067.

\bibitem{MN}
F. Magniez, A. Nayak.
Quantum complexity of testing group commutativity. 
{\em Proceedings of ICALP'05}, pp. 1312-1324.
Also quant-ph/0506265.

\bibitem{MSS}
F. Magniez, M. Santha, M. Szegedy. 
Quantum algorithms for the triangle problem. 
{\em Proceedings of SODA'05}, pp. 1109-1117.
Also quant-ph/0310134.

\end{thebibliography}
\end{document}